\documentclass[sigconf]{acmart}
\AtBeginDocument{%
  }

\copyrightyear{2026}
\acmYear{2026}
\setcopyright{cc}
\setcctype{by}
\acmConference[RecSys '26]{20th ACM Conference on Recommender Systems}{September 27-October 02, 2026}{Minneapolis, MN, USA}
\acmBooktitle{20th ACM Conference on Recommender Systems (RecSys '26), September 27-October 02, 2026, Minneapolis, MN, USA}
\acmDOI{10.1145/3773078.3831760}
\acmISBN{979-8-4007-2284-4/2026/09}

\usepackage{multirow}
\usepackage{enumitem}
\usepackage{array}
\usepackage{setspace}

\begin{document}

\title[Sparse Representations of Multimodal Content for Cold Item Recommendation]{Learning Sparse Representations of Multimodal Content for Enhanced Cold Item Recommendation}

\author{Gregor Meehan}
\email{gregor.meehan@qmul.ac.uk}
\orcid{0009-0007-2619-9299}
\affiliation{%
  \institution{Queen Mary University of London}
  \city{London}
  \country{United Kingdom}
}

\author{Johan Pauwels}
\email{j.pauwels@qmul.ac.uk}
\orcid{0000-0002-5805-7144}
\affiliation{%
  \institution{Queen Mary University of London}
  \city{London}
  \country{United Kingdom}
}

\renewcommand{\shortauthors}{Gregor Meehan and Johan Pauwels}

\begin{abstract}
The scale and rapid growth of item catalogs in modern digital platforms present significant challenges to recommender system (RS) practitioners. Most RSs use embedding similarity to predict user-item preferences, but embedding storage and low-latency retrieval are challenging in industry-scale catalogs. Furthermore, newly added items do not have corresponding embeddings and cannot be recommended effectively; previous works often tackle this item cold-start problem by generating cold item representations from auxiliary content, such as images or descriptive text, so that user preferences can be predicted without historical interactions. In this paper, we argue that sparse embeddings have notable advantages over standard dense vectors in this content-based cold-start paradigm. We describe how existing cold-start training regimes can be adapted for sparse representation learning, and build on insights from linear attention to design a pre-sparsification activation technique that induces sharpness and denoising effects in learned item-item similarities. We show that the resulting sparse embeddings achieve significant improvements in cold-start recommendation accuracy over dense embeddings at considerably lower storage costs, especially for users with multiple interests.  Through comprehensive experiments on four multimodal RS datasets, we also demonstrate the interpretability of sparse content embeddings and their robustness in the trade-off between size and accuracy.

\end{abstract}




\maketitle
\section{Introduction} \label{sec:intro}
Producing useful predictions for new items is a common hurdle in recommender systems (RSs): deep learning-based collaborative filtering (CF) relies on user-item interaction histories, so cannot handle `cold' items~\cite{schein2002methods}. Instead, auxiliary content features, such as text descriptions or product images, are often used to predict user interest in cold items~\cite{zhang2025cold}. Many works~\cite{van2013deep, chen2022generative, bai2023gorec, CLCRec, volkovs2017dropoutnet, tamm2026leveraging} use these features to estimate CF embeddings for new items, although recent research~\cite{meehan2025inherited} suggests that the semantic gap between content and CF spaces limits this approach's efficacy, and that directly leveraging content similarity for prediction can be more effective~\cite{meehan2026semco}. 

Another considerable challenge in real-world RS deployment is that of storing and accessing the huge vector tables required by embedding-based RSs in large-scale industrial applications~\cite{guo2021scalefreectr, xu2021agile}. Much research~\cite{li2024embedding} focuses on embedding compression in these industry-scale RSs, attempting to balance the trade-off between data efficiency and representation quality. 
Recent studies~\cite{kasalicky2025future, wen2025beyond, vanvcura2026efficient} suggest that \emph{sparse representations}~\cite{lee2006efficient, wright2010sparse, zhang2015survey} are a promising direction for reducing embedding storage costs while maintaining strong RS performance and providing increased interpretability. Unlike dense vectors, sparse embeddings contain only a small number of activated (non-zero) neurons, e.g.\ 16 out of 2048. Along with compact storage (e.g.\ in Compressed Sparse Row (CSR)~\cite{eisenstat1977yale} format), this approach also supports efficient inverted index retrieval~\cite{bruch2024efficient, bruch2025efficient}. Furthermore, dot products between two sparse vectors with $k$ non-zeros is $\mathcal{O}(k)$, meaning that sparse embeddings with low active dimensions can outperform dense embeddings in retrieval speed~\cite{wen2025beyond}. 

These works~\cite{kasalicky2025future, vanvcura2026efficient} explore sparse representations in CF models on warm items. In this paper, we instead investigate their application to cold items, examining how multimodal item features can be transformed into sparse embeddings well-suited to content-based recommendation. We argue that, beyond reducing size,  sparsity has other notable advantages over dense embeddings in content-based cold-start contexts, particularly for users with diverse interests. We design a sparsification pipeline to maximize these benefits, enhancing cold recommendations by emphasizing the most relevant items in a user's interaction history. Our contributions are as follows:

\begin{itemize}[leftmargin=*]
\item We describe how existing cold-start RS representation learning regimes (e.g. \cite{meehan2026semco}) can be adapted for sparse embeddings;
\item Through a detailed case study, we examine the limitations of dense embeddings in this context. We identify two desirable properties (namely, sharpness and denoising) in learned item-item similarities, and describe how they can be induced for sparse embedding similarities by pre-sparsification activation functions;
\item Through experiments on four multimodal RS datasets, we show that sparse embeddings consistently outperform dense methods by 16.6\% to 75.5\% in cold NDCG@20 (with comparable storage budgets), plus further gains on users with diverse interests;
\item We analyze other properties of sparse content embeddings, including their interpetability and robustness to level of sparsity.
\end{itemize} 
To facilitate reproducibility of our results, we release our code at \href{https://github.com/gmeehan96/SparseColdStart}{https://github.com/gmeehan96/SparseColdStart}.

\section{Related Work}
\subsection{Content-Based Cold-Start} \label{sec:rel_cold}
In this work, we address the strict item version of the cold-start problem, distinguishing it from recommendation of `long-tail' items with sparse but non-zero interaction histories. Some recent studies (e.g. \cite{wang2023zero,wu2024could,wu2024coral,huang2025large}) address item cold-start by leveraging the world knowledge of large language models (LLMs). Although promising, these methods face latency and efficiency challenges at industry scale~\cite{wang2024fresh,zhang2025cold} due to intensive LLM processing or retrieval-augmented generation at inference. For example, ColdLLM~\cite{huang2025large}, a recent work using LLMs to simulate cold item behavior, has a processing time of ten seconds per cold item. We focus on low-latency dot product retrieval, with inference times in the milliseconds. 

We therefore discuss methods that leverage item content to predict user preferences without relying on external corpora. DropoutNet~\cite{volkovs2017dropoutnet} simulates `coldness' during training by randomly replacing warm item CF embeddings with content features; recent methods augment this dropout paradigm with multi-task training~\cite{du2020learn}, mixtures-of-experts~\cite{zhu_recommendation_2020}, graph neighborhoods~\cite{NEMCF,kim2024content}, or contrastive alignment between ID and content embeddings~\cite{CLCRec, zhou2023contrastive, wang2024preference}.

However, forcing user embeddings to align with item content can harm warm item accuracy ~\cite{huang_aligning_2023}. Several works therefore use embeddings from a pre-trained warm RS to supervise the item content encoder, e.g.\ via mean squared error~\cite{van2013deep}, generative-adversarial nets~\cite{goodfellow2014generative, sun2020lara,chen2022generative} or variational autoencoders~\cite{Kingma2014,zhao2022improving,bai2023gorec}. Others~\cite{huang_aligning_2023, zhang2023dual} distill user-item scores from pre-trained warm RSs. Once trained, these models are used for new items, while the original CF model is applied to warm items, preserving warm performance.

Although more effective than dropout approaches, these generative methods have several drawbacks. First, they often require separate user embeddings for warm and cold items (e.g.~\cite{huang_aligning_2023, zhu_recommendation_2020, chen2022generative}), further increasing storage requirements. Recent work~\cite{meehan2025inherited} also shows that such methods inherit warm model predictive bias due to misalignment between CF and content signals. SEMCo~\cite{meehan2026semco} therefore frames content-based cold-start as an item similarity problem, using the structure of a shallow linear autoencoder~\cite{vanvcura2022scalable} to contrastively train multimodal content representations that correlate with user preferences. The resulting RS outperforms models reliant on CF embeddings, while also producing more equitable outcomes across the item population. We therefore adopt this content similarity paradigm for our sparse representation training regime.

\subsection{Sparse Representations}

Representation sparsity has been explored in information retrieval and question-answering contexts~\cite{formal2021splade, zamani2018neural, paria2020minimizing}, and sparse autoencoders (SAEs)~\cite{cunningham2023sparse, gao2024scaling} have recently gained attention for facilitating interpretability in LLMs. While earlier works encourage sparsity through regularization, such as penalties proportionate to L1 norms or numbers of FLOPS~\cite{paria2020minimizing}, we follow more recent studies (e.g.\ \cite{gao2024scaling, wen2025beyond, vanvcura2026efficient}) in enforcing sparsity explicitly via top-K activation, i.e.\ only keeping the top-K largest neurons in a dense embedding.

Recent work shows that applying SAEs to pre-trained dense embeddings can outperform other embedding compression techniques (e.g.\ Matryoshka~\cite{kusupati2022matryoshka}) in RSs~\cite{kasalicky2025future} and image retrieval~\cite{wen2025beyond}. Exact and approximate nearest neighbor search for sparse vectors are supported in PostgreSQL vector library \emph{pgvector}, and, in \cite{kasalicky2025future}, the authors perform online evaluation with a large-scale production RS ($\mathcal{O}(10^8)$ items), indicating that sparse embeddings are practicable in real-world systems. Other studies~\cite{vanvcura2026efficient}  train sparse embeddings from scratch for warm item CF, rather than as a post-hoc compression step; their sparse embeddings achieve comparable accuracy to dense vectors at a fraction of the storage cost, and facilitate segment interpretability in the learned user and item embeddings. In this work, we build on this approach to demonstrate the effectiveness of sparse embeddings in item cold-start contexts.

\section{Sparse Representations for Cold-Start}\label{sec:motivation}
We first outline the content similarity-based cold-start approach, then explain why sparse embeddings are suited to this paradigm.

\subsection{Preliminaries}\label{sec:prelim} We begin by revisiting SEMCo~\cite{meehan2026semco}, a training regime for content-based cold-start via contrastive learning on user-item interactions. Given binary user-item interaction matrix $\mathbf{R}\in\{0,1\}^{|\mathcal{U}|\times |\mathcal{I}^w|}$ storing historical interactions between user set $\mathcal{U}$ and warm item set $\mathcal{I}^w$, SEMCo learns an attention-based projection $f$ that fuses multimodal item content vectors $\mathbf{x}_1,...,\mathbf{x}_M$ into an embedding $\mathbf{y}$, i.e.
\begin{equation}
    \mathbf{y}=f(\mathbf{x}_1,...,\mathbf{x}_M)\in\mathbb{R}^d,
\end{equation}
before $\mathbf{y}$ is L2-normalized. For cold item set $\mathcal{I}^c$, user preference score matrix $\mathbf{P}\in\mathbb{R}^{|\mathcal{U}|\times |\mathcal{I}^c|}$ is predicted via matrix multiplication 
\begin{equation}
    \mathbf{P} = (\mathbf{R}\mathbf{Y})\mathbf{C}^\top,
\end{equation} 
where $\mathbf{Y}$ and $\mathbf{C}$ are the projected item content matrices for $\mathcal{I}^w$ and $\mathcal{I}^c$. In other words, users are represented by $\mathbf{R}\mathbf{Y}$, the aggregated embeddings of their interacted items, and SEMCo's contrastive objective ensures that similarity with $\mathbf{C}$ in the projected space correlates with user interest. This is a content-based adaptation of linear autoencoders (LAEs)~\cite{ning2011slim, steck2019embarrassingly}, which are often used in warm item CF; similarly to shallow LAEs~\cite{vanvcura2022scalable, vanvcura2025evaluating, vanvcura2024beeformer}, SEMCo factorizes the LAE item-item similarity matrix to facilitate inference on cold items. 

\subsection{Case Study}\label{sec:case_study}

Representing users by average embeddings of interacted items is a key efficiency in this factorized LAE setup. It avoids large $\mathcal{O}(|\mathcal{I}^w|^2)$ matrices during training, preserves linearity, and allows for user scores for new items to be computed by a single dot product. However, this approach also has notable downsides for accurate preference prediction. Users can have many diverse interests, so only a small subset of previously interacted items may be relevant to a target item; for example, users on music streaming platforms may not expect their suggestions for classical music to be affected by recently listened pop tracks. Multi-interest modeling is a well-studied challenge in the RS literature~\cite{Guo2021TheSP, cen2020controllable, li2019multi}, but is exacerbated by this user representation strategy, as unrelated interacted items can add significant noise to predicted preferences.

\subsubsection{Score Augmentation} 

\begin{figure*}
    \includegraphics[width=\linewidth,trim={0 0.85cm 0 0.0cm}]{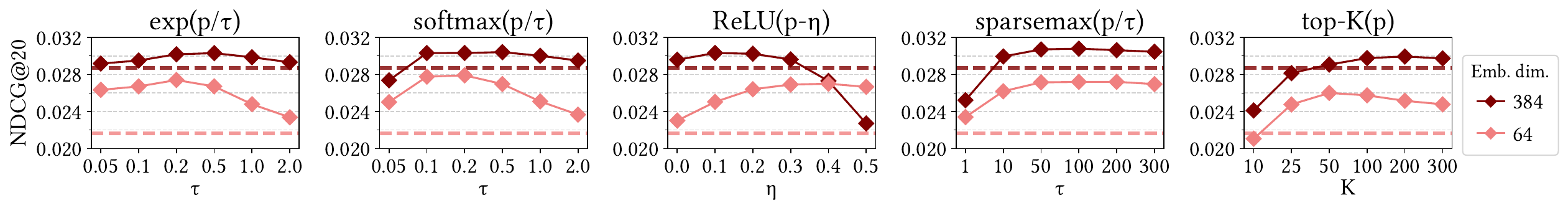} 
    \caption{Cold NDCG@20 on Electronics for functions $\phi$ of item-item similarities $p$ on SEMCo embeddings. Softmax, sparsemax, and top-K (i.e., KNN) are applied over warm items (columnwise on $\mathbf{Y}\mathbf{C}^\top$). The reference lines are SEMCo's base NDCG without $\phi$.}
    \label{fig:case_study_fns}
\end{figure*}

\begin{figure}
    \includegraphics[width=\linewidth,trim={0 0.85cm 0 0.7cm}]{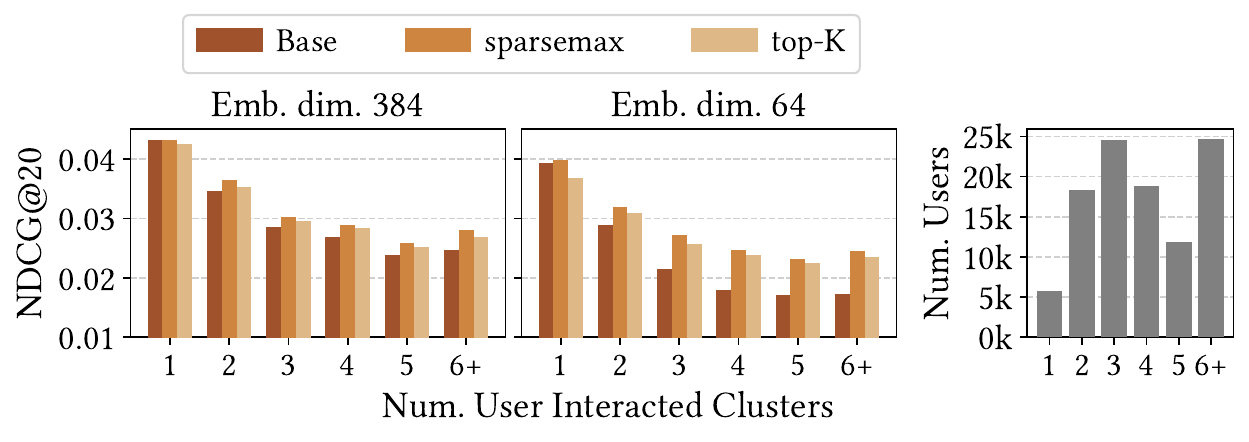} 
    \caption{Cold NDCG and count for users grouped by number of interacted clusters on Electronics.}
    \label{fig:case_study_user}
\end{figure}

To illustrate these limitations, we relax the linearity requirement in our preference predictions, and for a non-linear function $\phi$ we define augmented scores
\begin{equation}
    \mathbf{\tilde{P}} =\mathbf{R}\cdot\phi(\mathbf{Y}\mathbf{C}^\top). \label{eq:aug_pref}
\end{equation} 
There are two desirable properties of $\phi$ for addressing our concerns above. The first is \emph{denoising}, i.e.\ muting low or negative scores that can accumulate across large item sets to obscure useful items. The second is \emph{sharpening}, i.e.\ emphasizing the highest similarities so that they have the most influence on the predicted score; for lower scores, this is a softer version of denoising. To assess the impact of these properties, we train a SEMCo model on the Amazon Electronics dataset~\cite{mcauley2015image} following the experimental setup in~\cite{meehan2026semco}, and measure cold-start performance for different choices of $\phi$.

\subsubsection{Analysis}
We examine five $\phi$ functions, displaying results in Figure~\ref{fig:case_study_fns}: exp and softmax only sharpen; ReLU and sparsemax~\cite{martins2016softmax} both sharpen and denoise; and top-K only denoises. We note first that the larger embeddings (384d) provide much better preference modeling, with a gain in base NDCG@20 of 32.6\% over 64d vectors. The $\phi$'s consistently outperform the base embeddings, particularly at size 64, where they largely close the gap to 384d. In other words, filtering out noisy similarities compensates for the reduced semantic expressiveness of more compact embeddings. All $\phi$ with sharpening outperform top-K, i.e.\ accentuating high scores is beneficial beyond de-emphasizing less relevant items. Here $\tau$, $\eta$, and K control the strength of sharpening/denoising, and for each $\phi$ there is a balance between relevant scores being either under- or over-represented. 

To test our hypothesis about user interest diversity, we identify interests by dividing the Electronics items into 64 clusters via K-means~\cite{macqueen1967some} on the size 384 SEMCo embeddings. We group users by their number of interacted clusters and report the user count and average NDCG per group in Figure \ref{fig:case_study_user}. Results for base and augmented SEMCo predictions are similar for users with a single interest, and all methods see lower NDCG as the number of interests grows, with the difference again more drastic for the smaller 64d embeddings. However, the drop in accuracy is much smaller for the augmented predictions; since most users have at least three interests, dense embeddings have clear limitations for user preference modeling.

\subsection{Spikiness and Linear Attention}\label{sec:spikiness}

Denoising and sharpening both reduce the entropy of cold item similarity distributions across warm items. This `spikiness' in distribution has been shown to be a key property of effective attention weights in transformer-based language models~\cite{zhang2024hedgehog}. Indeed, when $\phi$ is softmax, Equation \ref{eq:aug_pref} can be viewed as an attention calculation: the queries are the cold item vectors $\mathbf{C}$, the keys are the warm item embeddings $\mathbf{Y}$, and the values are the interaction matrix $\mathbf{R}$. 

However, as with attention, non-linearity harms computational efficiency, as it requires materializing full similarity matrices. In RS cold-start, calculating cold item similarities to the full warm item set is impractical in industry-scale catalogs. While this can be mitigated through approximate methods such as FAISS~\cite{douze2024faiss} or Annoy~\cite{Github:annoy}, direct prediction by vector dot product is still preferable~\cite{covington2016deep}.  

Much research attempts to improve transformer efficiency by linearizing attention~\cite{zhang2024hedgehog, chou2024metala, chen2021scatterbrain, zhang2024lolcats, xiong2021nystromformer, katharopoulos2020transformers, choromanski2020rethinking}, often through low-rank approximations of the exp kernel. For self-attention over keys $\mathbf{K}$ and queries $\mathbf{Q}$ of length $n$, these methods apply some $\psi$ such that 
\begin{equation}
     \textrm{softmax}(\mathbf{K}\mathbf{Q}^\top) \approx \psi(\mathbf{K})\psi(\mathbf{Q})^\top,
\end{equation}
Then, for values $\mathbf{V}$, the attention calculation is linearized as 
\begin{equation}
     \mathbf{V}\cdot\textrm{softmax}(\mathbf{K}\mathbf{Q}^\top) \approx \mathbf{V}\cdot\psi(\mathbf{K})\psi(\mathbf{Q})^\top\\
     = (\mathbf{V}\cdot\psi(\mathbf{K}))\psi(\mathbf{Q})^\top,
\end{equation}
avoiding the $\mathcal{O}(n^2)$ cost of calculating the full $\mathbf{K}\mathbf{Q}^\top$ matrix. We take inspiration from this body of work: rather than applying a non-linearity $\phi$ to the item-item similarities, to preserve inference efficiency we want a non-linear transformation $\psi$ on the item content embeddings such that
$\psi(\mathbf{Y})\psi(\mathbf{C})^\top$ achieves the desired properties. 

We argue that embedding sparsification is a promising direction for defining $\psi$. Since dot products between dissimilar sparse vectors are typically zero, they naturally denoise without non-linear intervention. Below we describe how we adapt RS content representation learning regimes to produce sparse outputs, modifying standard sparsification pipelines to induce sharper similarity scores. 

\subsection{Learning Sparse Representations} 
Previous RS studies of sparse embeddings~\cite{kasalicky2025future, vanvcura2026efficient} focus on CF scenarios, where warm item ID vectors are learned from scratch. For content-based cold-start, our goal is instead to project item content vectors into a latent space that is useful for recommendation, i.e.\ where high cosine similarity corresponds to an increased likelihood of shared user interest. As described in Section \ref{sec:prelim}, content-based LAE-like methods (e.g.\ SEMCo~\cite{meehan2026semco}) perform well at this task with dense vectors; we therefore propose to modify their architecture to produce a sparse latent space with similar properties. We note that we only adjust the transformation pipeline compared to dense models, and do not modify the training objective. Similarly to~\cite{vanvcura2026efficient}, our sparse content representations are learned directly from interactions, rather than as an intermediate layer in an autoencoder~\cite{kasalicky2025future}.

\subsubsection{Sparsification} Content-based LAE methods learn a projection $f$ of multimodal features $\mathbf{x}_1,...,\mathbf{x}_M$ into a fused content representation $\mathbf{y}\in \mathbb{R}^d$. In SEMCo, $f$ is a two-layer single-branch MLP~\cite{ganhor2024multimodal} that adaptively fuses modalities with an attention mechanism. We follow recent work~\cite{vanvcura2026efficient, kasalicky2025future, wen2025beyond,gao2024scaling} and enforce sparsity in $\mathbf{y}$ by applying top-k activation for some $k$ before L2-normalization:
\begin{equation}
    \textrm{top-k}(\mathbf{y}[i]) = \begin{cases}
    \mathbf{y}[i] & \textrm{if } |\mathbf{y}[i]|\textrm{ is in the top }k\textrm{ elements of }  |\mathbf{y}|,\\
    0 & \text{otherwise} \\
\end{cases}
\end{equation}
where typically $k \ll d$. We keep the largest dimensions by absolute value, as in~\cite{vanvcura2026efficient}, to preserve top negative neurons. After sparsification, we L2-normalize and train the content LAE as normal.
 

\subsubsection{Activation}\label{sec:method_psaf} Since our embeddings are output by an MLP on content vectors, we can influence the learned latent space via additional transformations on $\mathbf{y}$. In particular, we aim to encourage sharper item-item similarities to complement the built-in denoising. We therefore introduce \textbf{p}re-\textbf{s}parsification \textbf{a}ctivation \textbf{f}unctions (\textbf{PSAFs}) to our sparse learning pipeline. This step is again motivated by techniques from linear attention, where several studies~\cite{xiong2021nystromformer, zhang2024hedgehog, zhang2024lolcats} find that applying exponential activations to query/key vectors is effective for increasing spikiness in attention matrices;  however, as with most linear attention methods, this typically requires larger embeddings to provide sufficient semantic flexibility for approximating the exp kernel.  By compressing this expressive capacity into very few active dimensions, sparse embeddings are ideal for a similar sharpening approach without this added cost.

Given our focus on embedding sparsity, we adopt the general $\alpha$-entmax function~\cite{peters2019sparse}, which interpolates between softmax ($\alpha=1$) and sparsemax ($\alpha=2$), as our PSAF. When $\alpha>1$, it is defined as 
\begin{equation}
\alpha\text{-entmax}(\mathbf{y})
=
\bigl[(\alpha - 1) \mathbf{y} - \eta \mathbf{1}\bigr]_+^{\frac{1}{\alpha - 1}},
\end{equation}
where $\mathbf{1}$ is the ones-vector, $\left[\cdot\right]_+$ is the positive-part operator, and $\eta$ is a threshold such that the output non-negative and sums to 1. Unlike softmax, where all outputs are positive, $\alpha$-entmax zeroes out low-scoring logits. As $\alpha$ grows, the level of sparsity increases, but the sharpness of the outputs decreases as the exponent term approaches 1. With appropriate tuning, $\alpha$-entmax thus provides flexibility for the content MLP to adaptively calibrate the level of sharpness and denoising for optimal user preference prediction.

\subsubsection{Two-Siding}\label{sec:method_ts} Since $\alpha$-entmax mutes negative neurons, we follow~\cite{zhang2024hedgehog, zhang2024lolcats, shang2016understanding} in preserving negative signals through antisymmetric concatenation before applying $\alpha$-entmax:
\begin{equation}
    \textrm{two-side}(\mathbf{y}) = [\mathbf{y};-\mathbf{y}]
\end{equation}
Although doubling embedding sizes during training, this step does not affect model parameters or storage post-sparsification, as we still keep only the top-k dimensions. We show in Section \ref{sec:results} that two-sided PSAFs can notably improve accuracy with only modest training cost. When using a PSAF, our $\psi$ function (Section \ref{sec:spikiness}) is
\begin{equation}
\psi(\mathbf{y}) =     \alpha\text{-entmax}(\textrm{two-side}(\mathbf{y})/\omega),
\end{equation}
where the hyperparameter $\omega$ modulates the sharpening effect.


\section{Experimental Design}
To validate the effectiveness of sparse representations for item cold-start, we investigate the following research questions:
\begin{itemize}[leftmargin=*]
\item \textbf{RQ1}: How do sparse embeddings perform compared to dense cold-start baselines, especially for users with multiple interests?
\item \textbf{RQ2}: What is the impact of pre-sparsification activation functions on item-item similarity distributions and training efficiency?
\item \textbf{RQ3}: How does varying the width and level of sparsity affect retrieval accuracy for sparse representations?
\item \textbf{RQ4}: To what extent do sparse embeddings facilitate interpretable content-based recommendations?
\end{itemize}

\subsection{Evaluation Protocol}
\begin{table}[]
    \caption{Dataset statistics. The feature vector dimension for each content modality is indicated in parentheses. \label{table:data_stats}}
  \small
  \setlength\extrarowheight{-1.25pt}
  \begin{tabular}{c@{\hskip 5pt}ccccc}
  \toprule
  Dataset                 & Users    & Items & Interactions & Density & Modes (dim)   \\ \midrule
  Clothing   & 39,387    & 23,033 & 278,677 & 0.031\% & Text (384) \\[1.5pt] 
  \multirow{2}{*}{Electronics} & \multirow{2}{*}{192,403}  & \multirow{2}{*}{63,001} & \multirow{2}{*}{1,689,188} & \multirow{2}{*}{0.014\%} & Image (4096) \\ 
  &                          &                         &                            &                          & Text (384) \\[1.5pt] 
     \multirow{2}{*}{M4A-Onion} & \multirow{2}{*}{8,807}  & \multirow{2}{*}{43,886} & \multirow{2}{*}{1,510,034} & \multirow{2}{*}{0.391\%} & Audio (1024) \\ 
  &                          &                         &                            &                          & Text (768) \\[1.5pt]  
  
  \multirow{3}{*}{Microlens} & \multirow{3}{*}{98,129}  & \multirow{3}{*}{17,228} & \multirow{3}{*}{705,174} & \multirow{3}{*}{0.042\%} & Image (1024) \\
  &                          &                         &                            &                          & Text (1024) \\
   &                          &                         &                            &                          &Video (768) 
  \\ \bottomrule
  \end{tabular}

  \end{table}

As summarized in Table~\ref{table:data_stats}, our datasets include two Amazon subsets~\cite{mcauley2015image} (namely, \textbf{Clothing} and \textbf{Electronics}) and \textbf{Microlens}~\cite{ni2023content}, with interaction data and preprocessed multimodal features available via MMRec~\cite{zhou2023mmrec}. We also test on the 2018 subset of Music4All-Onion~\cite{Music4All-Onion} (\textbf{M4A-Onion}) described in~\cite{meehan2026semco, meehan2025artist}, with MusicFM~\cite{won2024foundation} audio features and BERT~\cite{BERT} lyric features. 

To align with previous works in item cold-start~\cite{chen2022generative, huang_aligning_2023, meehan2025inherited, meehan2026semco}, we randomly select 20\% of items as our `cold' item set and divide this 50-50 into validation and test.\footnote{Although temporal splitting is a more realistic setting for production scenarios~\cite{meng2020exploring}, this approach can cause issues with sample sizes in cold item contexts~\cite{tamm2026leveraging}.} The remaining `warm' item interactions are split 80-10-10 into training, validation, and test sets for training the supervisory CF model for the cold-start baselines.    

We measure performance with standard ranking metrics Recall@$k$ and NDCG@$k$ in our cold holdout sets, setting $k=20$. All reported metrics are averaged over five runs at optimal hyperparameters. 

%

\begin{table*}[]
\caption{Cold test set Recall@20 (R@20) and NDCG@20 (N@20) for sparse embeddings and dense baselines. SEM-Sf and SEM-Sp stand for softmax and sparsemax SEMCo variants respectively. We \underline{underline} the best model in each group and \textbf{bold} the best overall. The improvement percentage is calculated for the best sparse method over the best dense content LAE method. 
}\label{tab:cold_results}
\small
\centering
\setlength\extrarowheight{-1.25pt}
\begin{tabular}{c@{\hskip4.0pt}c c@{\hskip4.0pt}c@{\hskip4.0pt}c@{\hskip4.0pt}c@{\hskip4.0pt}c      c@{\hskip4.0pt}c@{\hskip4.0pt}c c@{\hskip4.0pt}c@{\hskip4.0pt}c c@{\hskip4.0pt}c@{\hskip4.0pt}c@{\hskip6.0pt} c}
\toprule
       &  & \multicolumn{5}{c}{Dense (Baselines)} &       \multicolumn{3}{c}{Dense (Content LAEs)}     & \multicolumn{3}{c}{Sparse (PSAF: None)}  & \multicolumn{3}{c}{Sparse (PSAF: Sparsemax)} &                     
       \\ \cmidrule(l{3pt}r{3pt}){3-7} \cmidrule(l{3pt}r{3pt}){8-10} \cmidrule(l{3pt}r{3pt}){11-13} \cmidrule(l{3pt}r{4pt}){14-16}
Dataset                     & Metric  & ALDI   & CLCRec         & GAR    & GoRec        & Heater & ELSA  & SEM-Sf         & SEM-Sp      & ELSA  & SEM-Sf         & SEM-Sp         & ELSA  & SEM-Sf         & SEM-Sp          & \textit{Imp.}\\ \midrule
\multirow{2}{*}{Clothing}    & R@20   & 0.0954 & 0.1190         & 0.1148 & \underline{0.1389} & 0.1225 & 0.1459 & 0.1435          & \underline{0.1460} & \underline{0.1641} & 0.1544 & 0.1528 & 0.1666 & 0.1676          & \textbf{0.1711}   &17.2\%      \\
                             & N@20   & 0.0453 & 0.0599         & 0.0541 & \underline{0.0638} & 0.0561 & 0.0705 & 0.0698          & \underline{0.0716} & \underline{0.0822} & 0.0767 & 0.0763 & 0.0827 & 0.0824          & \textbf{0.0835}   &16.6\%   \\[2pt]
\multirow{2}{*}{Electronics} & R@20   & 0.0297 & 0.0299         & 0.0321 & \underline{0.0376} & 0.0329 & 0.0487 & 0.0482          & \underline{0.0488} & \underline{0.0650} & 0.0564 & 0.0550 & 0.0573 & 0.0631          & \textbf{0.0659}   &35.1\%    \\
                             & N@20   & 0.0130 & 0.0138         & 0.0148 & \underline{0.0171} & 0.0149 & 0.0217 & 0.0217          & \underline{0.0218} & \underline{0.0294} & 0.0258 & 0.0254 & 0.0265 & 0.0287          & \textbf{0.0302}   &38.5\%    \\[2pt]
\multirow{2}{*}{M4A-Onion}   & R@20   & 0.0406 & \underline{0.0577}   & 0.0464 & 0.0299       & 0.0544 & 0.0784 & 0.0798          & \underline{0.0808} & \underline{0.1026} & 0.0925 & 0.0892 & 0.1150 & 0.1201          & \textbf{0.1270}   &57.1\%  \\
                             & N@20   & 0.0403 & \underline{0.0557}   & 0.0467 & 0.0316       & 0.0557 & 0.0762 & 0.0765          & \underline{0.0769} & \underline{0.1049} & 0.0934 & 0.0944 & 0.1221 & 0.1212          & \textbf{0.1350}   &75.5\%    \\[2pt] 
\multirow{2}{*}{Microlens}   & R@20   & 0.1075 & 0.0914         & 0.1180 & \underline{0.1294} & 0.1203 & 0.1361 & \underline{0.1389} & 0.1357          & \textbf{0.1544} & 0.1456 & 0.1423 & 0.1460 & \underline{0.1483} & 0.1397            &11.1\%      \\
                             & N@20   & 0.0445 & 0.0401         & 0.0487 & \underline{0.0562} & 0.0504 & 0.0600 & \underline{0.0609} & 0.0591          & \textbf{0.0742} & 0.0671 & 0.0644 & 0.0666 & \underline{0.0694} & 0.0633            &21.7\%           \\
 \bottomrule        
\end{tabular}
\end{table*}

\subsection{Methods} 
In this section, we describe the training setup for our sparse representations and dense baseline methods. All models are trained with the Adam optimizer~\cite{Kingma2014AdamAM} and implemented using PyTorch~\cite{paszke2019pytorch}. 

\subsubsection{Sparse}
We test two content-based training regimes for our sparse representations. The first is the base \textbf{SEMCo}~\cite{meehan2026semco} model; unless otherwise specified, we use the sparsemax variant. The other is a content-based analogue of shallow LAE method \textbf{ELSA}~\cite{vanvcura2022scalable}, where the learned CF embeddings are replaced with the same attention-based multimodal content encoder as in SEMCo. To avoid dead latents~\cite{gao2024scaling} in our sparse embeddings, we use the exponential pruning strategy from~\cite{vanvcura2026efficient}, decaying the number of active dimensions from the full embedding width to 32 over the course of training. Unless otherwise stated, we use the top-32 dimensions at inference and set the embedding width of our content encoder output at 1024; for two-sided methods (see Section \ref{sec:method_ts}), the effective width is therefore 2048. The content encoder's hidden dimension is 384.

The search ranges for SEMCo's hyperparameters and the PSAF $\omega$ are on our GitHub. For ELSA, the learning rate is 0.0001 with early stopping on cold NDCG@20. We tune L2 regularization and batch size in $\{0,1\mathrm{e}^{-6},1\mathrm{e}^{-5},1\mathrm{e}^{-4}\}$ and $\{512,1024\}$. The SEMCo batch size is 2048, and we apply cosine learning rate decay~\cite{loshchilov2016sgdr} from 0.001 to zero over 40 epochs. We set the L2 regularization weight at $1\mathrm{e}^{-5}$ and $1\mathrm{e}^{-3}$ for the sparsemax and softmax SEMCo variants.

\subsubsection{Dense} 
Along with dense versions of SEMCo and ELSA, we implement five dense embedding cold-start baselines: 
\begin{itemize}[leftmargin=5mm]
    \item \textbf{CLCRec}~\cite{CLCRec} aligns CF and content features with a contrastive objective for improved cold item performance;
    \item \textbf{Heater} \cite{zhu_recommendation_2020} applies a dropout strategy~\cite{volkovs2017dropoutnet} during training and uses mixtures-of-experts \cite{shazeer2017outrageously} for item content transformation;
    \item \textbf{GAR}~\cite{chen2022generative} uses a generative-adversarial approach between the content-based generator and pre-trained warm model;
    \item \textbf{GoRec}~\cite{bai2023gorec} implements a conditional variational autoencoder~\cite{sohn2015learning} to generate item CF embeddings conditioned on content;
    \item \textbf{ALDI}~\cite{huang_aligning_2023} applies knowledge distillation to align cold student-warm teacher rating distributions and rankings.
\end{itemize}
We use multimodal RS FREEDOM~\cite{zhou_tale_2023} as the supervisory warm model for the generative methods (i.e., all except CLCRec), following existing work~\cite{meehan2026semco, meehan2025inherited}. In alignment with previous cold-start studies~\cite{meehan2026semco, meehan2025inherited, bai2023gorec, chen2022generative}, we set the user and item embedding dimension for our baseline methods at 64, and tune their other hyperparameters according to the ranges described in the original papers. For dense SEMCo and ELSA, we use the same hyperparameter search ranges as in the sparse case. For SEMCo, we additionally tune the L2 regularization in $\{1\mathrm{e}^{-5}, 1\mathrm{e}^{-4}, 1\mathrm{e}^{-3}, 1\mathrm{e}^{-2}\}$ and again use cosine learning rate decay from 0.001 to zero, over 15 epochs as in~\cite{meehan2026semco}.

\section{Results}\label{sec:results}

\subsection{Sparse vs. Dense (RQ1)} \label{sec:overall_results}

To measure the gains offered by sparse representations, we first evaluate them against our dense cold-start baselines with comparable storage budgets. Sparse vector storage requires saving both values and indices, so with $n$ active dimensions the storage cost is equivalent to that of a dense embedding of size $2n$.\footnote{We assume that the dense embeddings and sparse values are stored at the same precision. With full-precision floats, sparse embeddings are more space efficient because small integer indices use fewer bytes. For example, \texttt{float32} 64d dense embeddings on Electronics (63,001 items) use 16.1 MB; CSR sparse embeddings with width 1024, 32 active dimensions, index precision \texttt{int16}, and value precision \texttt{float32} take up 12.4 MB. If we use half-precision (\texttt{float16}) instead, the sizes are 8.1 MB and 7.3 MB for dense and sparse respectively. All experiments in this paper use \texttt{float32}.} We therefore set the number of active dimensions in the sparse representations at 32, since the embedding dimension of our dense baselines is 64. 

We note that, if item embeddings are sparsified,  user vectors can have many more than 32 active dimensions, as their item embeddings may contain diverse sets of non-zero neurons. We therefore also sparsify the user vectors to 32 active neurons to maintain a fair comparison; we quantify the impact of this step in Section \ref{sec:user_sparse}.  

\begin{figure*}
    \includegraphics[width=\linewidth,trim={0 0.85cm 0 0.0cm}]{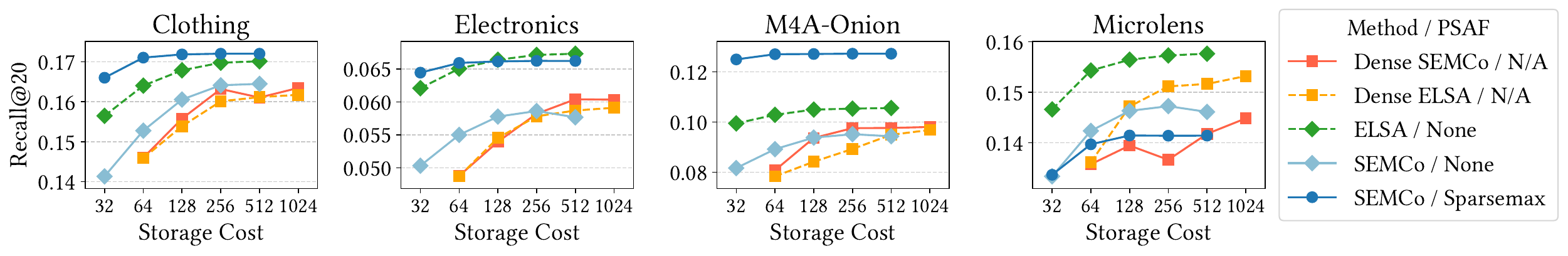} 
    \caption{Cold Recall@20 for dense embeddings against sparse embeddings with comparable storage cost (e.g.\ a 128-dimensional dense vector and sparse embedding with 64 active dimensions both have storage cost 128).}
    \label{fig:active_vs_dense}
\end{figure*}

\begin{figure*}
    \includegraphics[width=\linewidth,trim={0 0.85cm 0 0.0cm}]{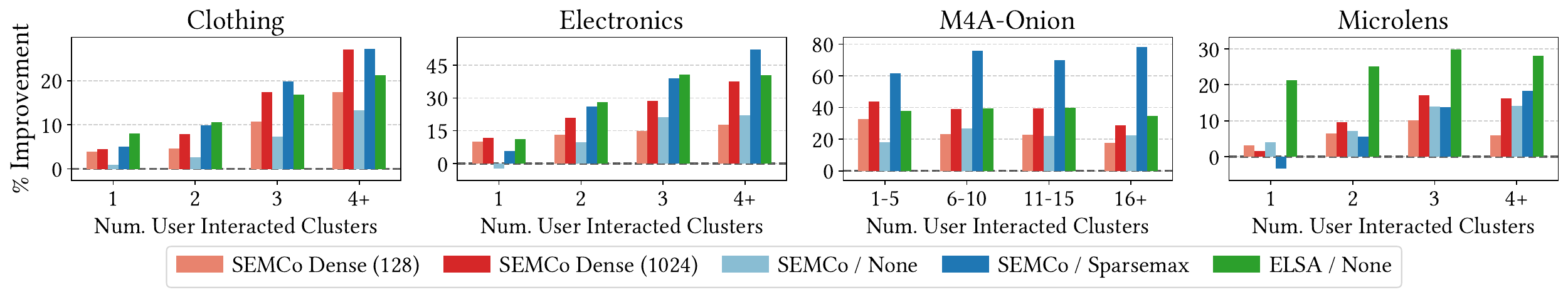} 
    \caption{Improvements in NDCG@20 over dense SEMCo with embedding size 64, against number of interacted clusters by user.}
    \label{fig:user_groups}
\end{figure*}

\subsubsection{Baseline Comparison} Table \ref{tab:cold_results} displays cold test set results for our dense and sparse methods. The sparse regimes consistently outperform both the CF-based baselines and the dense content LAEs: all gains of sparse methods over the corresponding dense LAE are statistically significant in both metrics by paired t-test with $p<0.05$ (except for R@20 on Microlens for Sem-Sp/Sparsemax) and with $p<0.005$ in all datasets except Microlens. In other words, sparsifying embeddings universally improves cold-start outcomes compared to dense representations with equivalent training regimes. 

The sparsemax variant of SEMCo with sparsemax PSAF has the best overall accuracy, again with the exception of Microlens, where ELSA without PSAF is strongest; we discuss Microlens's unique results in Section \ref{sec:res_microlens}. The sparsemax PSAF generally benefits SEMCo, but has more mixed outcomes for ELSA. We hypothesize that SEMCo's contrastive regime is better-suited for tuning the sharpening effect, as its extra $\alpha$-entmax on user-item scores per batch can adaptively modulate logit distributions; by contrast, ELSA reconstructs user-item scores directly with mean-squared error, which may harm training stability when the entropy of the MLP outputs is fluctuating. We analyze PSAFs further in Section \ref{sec:PSAF_results}.

\subsubsection{Storage Ablation} We measure sparse and dense embedding performance for varying storage costs in Figure \ref{fig:active_vs_dense}. We use fixed sparse embeddings with varying numbers of active dimensions, while the dense model hyperparameters are re-tuned at each embedding size to maximize performance. We cap the dimension at 1024, since most settings see diminishing returns beyond this size.

Existing work~\cite{vanvcura2026efficient} shows that, in warm item CF, sparse representations achieve broadly equivalent accuracy to dense models but with significantly lower storage requirements. In our content-based cold-start setting, we see  that sparse embeddings are superior on both counts: in all datasets, at least one sparse method with storage cost 64 outperforms the 1024d dense embeddings, representing 16$\times$ storage savings with improved accuracy. The best sparse embeddings with storage cost 128 improve over 1024d dense SEMCo embeddings by at least 5.2\% (Clothing) and up to 29.7\% (M4A-Onion). 

Other than reducing storage, previous research~\cite{wen2025beyond} shows that lowering active dimensions materially increases retrieval speed of sparse embeddings, making them competitive with or faster than dense vectors, especially in large databases. Given these benefits to practicality and accuracy, sparse representations offer compelling advantages over dense embeddings for content-based cold-start.

\begin{figure}
     \includegraphics[width=\linewidth,trim={0.0cm 0.95cm 0.0cm 0.5cm}]{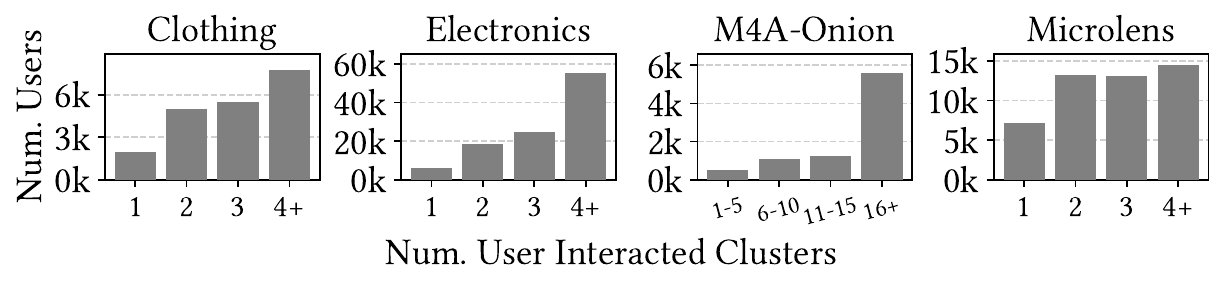}
    \caption{Distributions of number of user interacted clusters.}
    \label{fig:cluster_counts}
\end{figure}

\subsubsection{Interest Diversity}\label{sec:res_discussion} To understand why sparse embeddings consistently outperform dense embeddings, we recall the insights from our case study in Section \ref{sec:case_study}. In Figure \ref{fig:user_groups}, we visualize gains over 64d SEMCo embeddings by user interest diversity; we determine interests as described in Section \ref{sec:case_study}, namely by KMeans with 64 groups on SEMCo embeddings, with user counts by group displayed in Figure~\ref{fig:cluster_counts}. Due to the high interaction density of M4A-Onion, we use wider bins to maintain an informative distribution.  

As we saw with the $\phi$ non-linearities, the sparse methods achieve increasing gains over the dense baseline as the number of user interests grows.  Furthermore, for users with at least two interests, SEMCo/Sparsemax generally outperforms the 1024d dense embeddings, while SEMCo/None lags behind. This result illustrates that the PSAF sharpening is particularly valuable for modeling users with multiple interests in the SEMCo setting, as hypothesized above. This is highlighted by SEMCo/Sparsemax's exceptional performance in M4A-Onion, where users have much higher numbers of interests, meaning it is particularly important to emphasize relevant items when determining preferences. In general, the overall gains in each dataset in Table \ref{tab:cold_results} correlate exactly with the level of positive skew in interest count distribution in Figure~\ref{fig:cluster_counts}.

\subsubsection{Microlens}\label{sec:res_microlens} We next briefly discuss Microlens's disparate behavior from other datasets. It has relatively low gains over dense baselines, especially for SEMCo, and furthermore is the only dataset where the sparsemax PSAF does not improve the SEMCo sparsemax variant. We attribute this in part to Microlens's unique interaction structure, with almost 6$\times$ as many users as items and the lowest proportion of users with multiple interests (see Figure~\ref{fig:cluster_counts}), where the gains from sparsity and sharpening are largest. 

Based on its superior performance, the reconstruction training objective of ELSA is also clearly better-suited to this interaction structure than SEMCo's contrastive regime; ELSA performs well on Microlens even without sparsification, as can be seen in Figure~\ref{fig:active_vs_dense}. Based on our findings, we suggest to practitioners that SEMCo/Sparsemax is optimal for applications with high user activity, while ELSA without PSAF is also a consistently good choice, especially for smaller item catalogs or less active userbases.

\begin{table}[]

\setlength\extrarowheight{-1.25pt}
\caption{Cold NDCG@20 for PSAF ablation in SEMCo. \label{table:psaf}}
\small
\begin{tabular}{llc@{\hskip6.0pt}c@{\hskip6.0pt}c@{\hskip6.0pt}c}
\toprule
                           & PSAF       & Clothing & Elec. & M4A-Onion & Microlens \\ \midrule
\multirow{2}{*}{One-Sided} & None       & 0.0763   & 0.0254      & 0.0944    & \underline{0.0644}    \\
                           & Sparsemax  & 0.0819   & 0.0267      & 0.1224    & 0.0574    \\[2pt]
\multirow{4}{*}{Two-Sided} & ReLU       & 0.0797   & 0.0268      & 0.1014    & 0.0635    \\
                           & Softmax    & 0.0778   & 0.0263      & 0.1135    & \textbf{0.0668}    \\
                           & 1.5-Entmax & \underline{0.0832}   & \underline{0.0285}      & \textbf{0.1385}    & 0.0557    \\
                           & Sparsemax  & \textbf{0.0835}   & \textbf{0.0302}      & \underline{0.1350}    & 0.0633    \\
                           \bottomrule
\end{tabular}
\end{table}

\subsection{PSAFs (RQ2)}\label{sec:PSAF_results}

We next examine the impact of our pre-sparsification activations on model behavior, reporting SEMCo performance for various PSAFs in Table \ref{table:psaf}. Aside from $\alpha$-entmax, we also include CReLU~\cite{shang2016understanding}, or two-sided ReLU, which zeroes out the redundant negatives after two-siding. We see that this generally improves accuracy compared to only sparsification, as the wider embeddings facilitate denoising. Its zeroing effect on negative similarities is also beneficial, as seen in the ReLU ($\eta=0$) results in Figure~\ref{fig:case_study_fns}. Sparsemax also performs better after the two-siding operation, illustrating the value of preserving negative MLP outputs. Overall, the two-sided $\alpha$-entmax variants consistently have the highest accuracy across all datasets (again with the exception of Microlens), especially for $\alpha$ at 1.5 or 2.

\subsubsection{Sharpness and Denoising}
\begin{figure}
     \includegraphics[width=\linewidth,trim={0.0cm 0.85cm 0.0cm 0}]{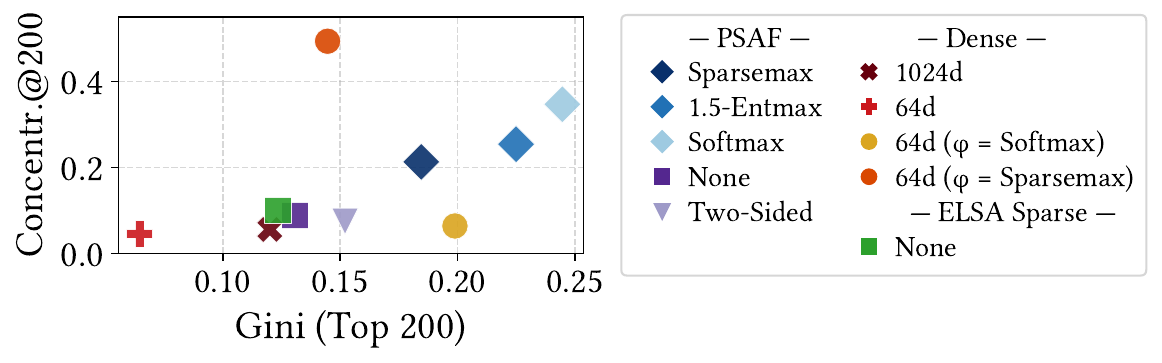}
    \caption{Gini index and concentration of top-200 cold-warm item similarities on Electronics for sparse SEMCo and others.}
    \label{fig:entropy}
\end{figure}

In Figure~\ref{fig:entropy}, we quantify the denoising and sharpening of our PSAFs on item-item similarity distributions. For cold items in Electronics, we take the top-200 warm item similarity scores and report two quantities: as a proxy for denoising, we calculate the \emph{concentration}, or proportion of total (non-negative) scores across the full warm item population made up by the top-200 items. To measure sharpness, we calculate inequality within those top-200 scores with the \emph{Gini index}, where values of 0 and 1 indicate maximal equality and inequality respectively. We use 200 because this provided the best top-K performance in Figure~\ref{fig:case_study_fns}.

We make several observations: first, as expected, 64d dense embeddings have minimal sharpness and denoising. Increasing the embedding size to 1024 notably improves the Gini index and slightly removes noise, and sparse embeddings at this size without PSAF only provide a very slight increase over the dense 1024d vectors.  Two-siding induces a further gain in Gini index, but the impact of the $\alpha$-entmax PSAFs is much clearer: they all materially increase both sharpness and denoising, with the degree of increase growing with $\alpha$ as hypothesized in Section \ref{sec:method_psaf}. For $\alpha < 2$, the sharpening impact is more pronounced even than the post-hoc application of softmax or sparsemax as $\phi$ to the 64d embedding similarities. 

We note that stronger sharpness and denoising effects do not always lead to improved performance, e.g. ELSA without PSAF also has generally strong accuracy. We hypothesize that sparsemax is generally the best-performing PSAF because it has the least extreme sharpening effect while still making a material impact, allowing for more stable convergence during SEMCo training. Given our results in Section \ref{sec:overall_results}, endowing the model with the ability to calibrate sharpening and/or denoising clearly enhances learning of sparse representations for cold item recommendation in most contexts.

\subsubsection{Training Efficiency}\label{sec:psaf_efficiency}
\begin{table}[]
\small
\caption{Training efficiency for dense and sparse SEMCo on Electronics. Soft. stands for softmax. \label{table:efficiency}}
\setlength\extrarowheight{-1.25pt}
\renewcommand{\arraystretch}{0.9}

\begin{tabular}{l@{\hskip 4.7pt}c c@{\hskip4.0pt}c c@{\hskip4.0pt}c@{\hskip4.0pt}c}
\toprule
               &   Dense    & \multicolumn{2}{c}{Sparse (One-Sided)} & \multicolumn{3}{c}{Sparse (Two-Sided)}  \\
              \cmidrule(l{1pt}r{5pt}){2-2} \cmidrule(l{3pt}r{3pt}){3-4} \cmidrule(l{3pt}r{3pt}){5-7}
PSAF               & N/A & None  & Sparsemax  & ReLU & Soft. & Sparsemax            \\ \midrule
GPU mem. (GB)  & 2.20  & 2.20  & 2.20       & 2.45 & 2.55    & 2.66  \\
Per epoch (sec) & 12.8 & 13.7 & 14.0       & 14.9 & 15.2    & 15.7   \\        \bottomrule
\end{tabular}
\end{table}

Although PSAFs generally increase accuracy, they require additional computation. In Table \ref{table:efficiency}, we quantify the cost of this extra activation step for SEMCo on Electronics, our largest dataset. We run all configurations on an NVIDIA H100 GPU. 

We note first that all sparse methods have slightly longer training time per epoch than dense 1024d embeddings due to the sparsification step, and that the two-siding operation also increases overall training cost. Since we implement sparsemax with efficient triton kernels~\cite{gonccalves2025adasplash}, it only leads to a moderate increase in cost despite being $\mathcal{O}(d \log d)$; however, the magnitude of this increase is higher for the two-sided embeddings, suggesting its effect may become slightly more pronounced for higher embedding widths. Overall, however, the memory and training costs of the best-performing sparse embeddings are only around 15\% to 20\%  higher than for dense embeddings or sparse embeddings without a PSAF. We argue that this modest increase in training cost does not outweigh the significant benefits to accuracy and storage discussed above.

\subsection{Embedding Properties (RQ3)}
In Figure \ref{fig:active_vs_dense}, we investigated sparse embedding quality for varying numbers of active dimensions. Below we perform further analysis of how embedding size and sparsity affect ranking accuracy.

\subsubsection{Embedding Width} 

\begin{figure}
     \includegraphics[width=\linewidth,trim={0.0cm 0.8cm 0.0cm 0}]{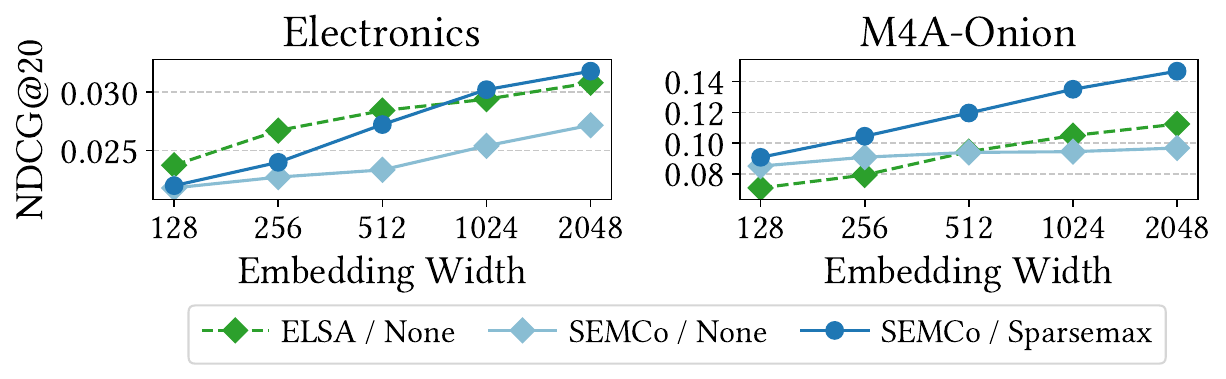}
    \caption{Cold NDCG@20 for sparse embeddings of varying widths with 32 active dimensions}
    \label{fig:width}
\end{figure}
Rather than varying the number of active neurons, we can also change the total available dimensions in our sparse embeddings. We note that, for consistency in parameter counts, where applicable we measure the embedding width before the two-side operation, i.e.\ as the dimension of the output of the content encoder MLP. In Figure \ref{fig:width}, we see that using wider embeddings generally improves accuracy at a logarithmic rate in Electronics and M4A-Onion; trends in our other datasets are similar. We also observe that, for small widths, performance is similar for SEMCo with and without sparsemax PSAF, but that the margin widens as the width increases. This again illustrates that including the PSAF allows SEMCo to use the embedding space more effectively.

Although improving ranking accuracy, larger embedding widths can lead to increased training costs, as seen in our comparison between one- and two-sided methods in Section \ref{sec:psaf_efficiency}. We therefore fix embedding widths at 1024 (or 2048 for two-sided methods) in our other experiments, as this provides consistently strong performance at moderate training speed in our offline scenario. However, previous studies~\cite{wen2025beyond} show that wider embeddings have faster retrieval, as they have fewer index overlaps and therefore require fewer multiplication operations. Although this balance between training cost and inference efficiency may require calibration in real-world applications, our results show that sparse embedding retrieval performance is generally robust to these adjustments.

\subsubsection{User Sparsification}\label{sec:user_sparse} 
\begin{figure}
     \includegraphics[width=\linewidth,trim={0.0cm 0.85cm 0.0cm 0}]{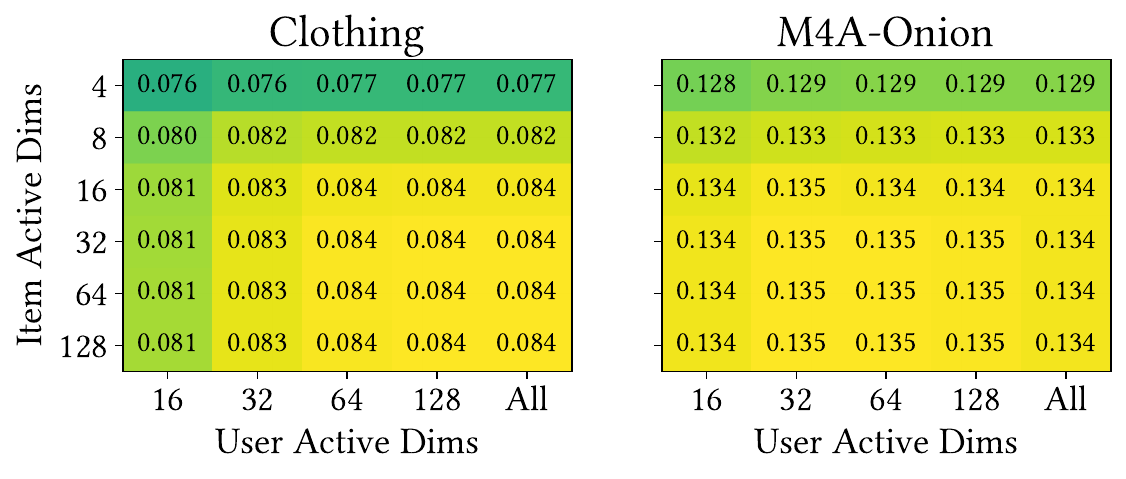}
    \caption{Cold NDCG@20 over varying user and item active dimensions for SEMCo with sparsemax PSAF.}
    \label{fig:user_item_active}
\end{figure}

As noted in Section \ref{sec:overall_results}, to maintain a fair storage comparison with dense vectors, we sparsify the user embeddings to the same number of active dimensions as the item embeddings. In theory, this can remove important user preference signal, especially for highly active users with diverse interests. We therefore measure the cost of this step in Figure~\ref{fig:user_item_active}; we display results for Clothing and M4A-Onion, but see similar trends for the other datasets. Accuracy is in fact largely robust to this user sparsification: there is little benefit to keeping the number of active user dimensions higher than that of items, providing comfort that pruning user embeddings does not overly damage ranking performance.

\subsection{Interpretability (RQ4)}
Beyond their considerable performance gains, sparse embeddings can also facilitate increased interpretability of the learned representations~\cite{cunningham2023sparse}. Previous studies~\cite{vanvcura2026efficient} show that sparse latent factors in CF models form semantically coherent groups, which allows for segment-based analysis of user behavior. We extend this idea to content-based representations with a case study on M4A-Onion, using the top 100 genre labels (by frequency among tracks) from Music4All~\cite{santana2020music4all} as our `ground-truth' for measuring semantic consistency. For tracks with more than one label, we keep the least popular genre as the single ground-truth value.


As seen in Figure \ref{fig:user_item_active}, accuracy in M4A-Onion is robust to high sparsity. We therefore activate the top-8 dimensions per warm item in SEMCo with sparsemax PSAF and measure pointwise mutual information (PMI) between binarized dimension activation and genres. For each dimension, we refer to the label with maximum PMI as its \emph{primary correlate}. The average primary correlate PMI across dimensions is 3.81, i.e.\ a dimension being active makes an item belonging to its primary correlate genre $2^{3.81}\approx14\times$ more likely than chance, indicating strong correlation between dimensions and individual labels. Furthermore, all 100 genres are the primary correlate of at least one dimension, showing that the learned embeddings provide strong coverage of ground-truth genres. This disentanglement illustrates how sparse embeddings better model diverse user interests by reducing noisy collisions among dissimilar items. 

\begin{figure}
     \includegraphics[width=\linewidth,trim={0.0cm 0.95cm 0.0cm 0}]{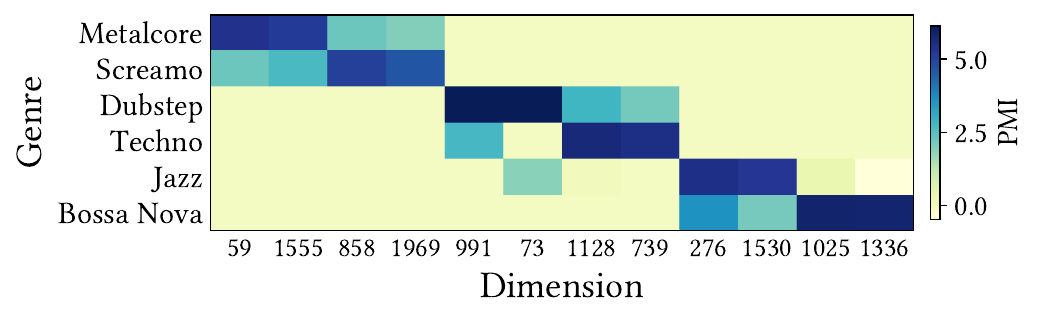}
    \caption{Top-2 correlated dims by PMI for selected genres.}
    \label{fig:genres_pmi}
\end{figure}

In Figure \ref{fig:genres_pmi}, we visualize this correlation for three pairs of related genres, namely Metalcore/Screamo, Dubstep/Techno, and Jazz/Bossa Nova. We see a clear block diagonal structure: the PMI is typically zero for dissimilar styles, while each label has its own specialized dimensions and moderate correlation  for related genres. 
It is thus directly observable through analysis of embedding activation patterns that the sparse embeddings capture our intuition about the relationship between genre similarity and user preference.  

This analysis shows that the disentanglement of user interests in our content-based sparse embeddings provides semantic interpretability alongside strong predictive accuracy.  If semantic labels are available, as for M4A-Onion, then we can use these directly to interpret user interests. However, in real-world catalogs, such labels are often missing or of poor-quality~\cite{sordo2008quest}, particularly for newly added items. By contrast, our sparse latent dimensions are learned only from item content (in this case, audio and lyric embeddings) and user interactions, and do not require supervision from metadata labels. Similarly to ~\cite{vanvcura2026efficient}, the resulting latents can be therefore be used to derive coherent item segments that facilitate interpretability of user interests. We have shown that similar insights apply to our sparse embeddings of multimodal item content, opening new avenues for interpretable content-based recommendation.

\section{Conclusion}
In this paper, we present a pipeline for learning sparse representations of multimodal content in RS item cold-start contexts. Through comprehensive experiments on four datasets, we show that sparse embeddings consistently and significantly outperform dense methods with equivalent training regimes and storage budgets in cold recommendation accuracy, especially for users with diverse interests. Furthermore, we describe how item-item similarities in the learned latent space can be manipulated via activation functions to exhibit desirable sharpness and denoising properties. Together with their robustness to sparsity level and their increased interpretability, our results show that sparse embeddings are a compelling alternative to standard dense vectors for content-based item cold-start. Future work will investigate how these benefits can be expanded to enhance content-based recommendation more generally, e.g.\ for missing modality scenarios~\cite{ganhor2024multimodal, malitesta2026training, bai2024multimodality} or to improve outcomes for long-tail items in warm multimodal RSs. 

%

\begin{acks}
This work was funded by UKRI as part of the UKRI CDT in Artificial Intelligence and Music [grant number EP/S022694/1].
\end{acks}

\bibliographystyle{ACM-Reference-Format}
\bibliography{bibfile}


\end{document}